\documentclass[preprint,review]{elsarticle}

\usepackage[margin=1.35cm]{geometry}
\usepackage{setspace}

\usepackage{lineno}
\usepackage{xcolor}
\usepackage{url}
\usepackage[hyperindex,breaklinks]{hyperref}
\hypersetup{linkcolor=red, citecolor=red, colorlinks=true}
\usepackage{amsmath}
\usepackage{amsfonts}
\usepackage{amssymb}
\usepackage{amsthm}

\usepackage{graphicx}
\usepackage[capitalise]{cleveref}
\usepackage[caption = false]{subfig}
\usepackage{placeins}

\journal{Physica A}

\usepackage{soul}

\usepackage{graphicx}
\usepackage{dcolumn}
\usepackage{bm}
\usepackage{hyperref}

\usepackage{url}
\usepackage[hyperindex,breaklinks]{hyperref}
\hypersetup{linkcolor=blue, citecolor=red, colorlinks=true}

\usepackage{xcolor}
\usepackage{xparse}

\begin{document}

\begin{frontmatter}

\title{
Cryptocurrency Time Series on the Binary Complexity-Entropy Plane:\\ Ranking Efficiency from the Perspective of Complex Systems
}

\author[add1]{Erveton P. Pinto}

\author[add2,add4]{Marcelo A. Pires}
\ead{piresma@cbpf.br, corresponding author.}

\author[add3]{Rone N. da Silva}

\author[add4,add5]{S\'\i lvio M. Duarte~Queir\'{o}s}

\address[add1]{Departamento de Ciências Exatas e Tecnológicas, Universidade Federal do Amapá
, Macapá, Amapá 68903-419, Brazil}

\address[add2]{Eixo de Tecnologia, Universidade Federal de Alagoas - Campus do Sertão, Delmiro Gouveia, Alagoas 57480-000, Brazil}

\address[add3]{Secretaria Municipal de Educação de Gurupá,  Gurupá - PA, 68300000, Brazil}

\address[add4]{Centro Brasileiro de Pesquisas F\'{\i}sicas, Rio de Janeiro - RJ, 22290-180, Brazil}

\address[add5]{National Institute of Science and Technology for Complex Systems, Brazil}

\begin{abstract}
We report the first application of a tailored Complexity-Entropy Plane designed for binary sequences and structures. We do so by considering the daily up/down price fluctuations of the largest cryptocurrencies in terms of capitalization (stable-coins excluded) that are worth $circa \,\, 90 \%$ of the total crypto market capitalization. With that, we focus on the basic elements of price motion that compare with the random walk backbone features associated with mathematical properties of the Efficient Market Hypothesis. 
From the location of each crypto on the Binary Complexity-Plane (BiCEP) we define an inefficiency score, $\mathcal I$, and rank them accordingly.
The results based on the BiCEP analysis, which we substantiate with statistical testing, indicate that only Shiba Inu (SHIB) is significantly inefficient, whereas the largest stake of crypto trading is reckoned to operate in close-to-efficient conditions. Generically, our $\mathcal I$-based ranking hints the design and consensus architecture of a crypto is at least as relevant to efficiency as the features that are usually taken into account in the appraisal of the efficiency of financial instruments, namely canonical fiat money. Lastly, this set of results supports the validity of the binary complexity analysis.
\end{abstract}

                              
\end{frontmatter}


\section{Introduction}

Although the roots of financial efficiency were already put forward as early as the 1560s by Cardano within a gambling context~\cite{cardano}, it is attributed to Fama and Malkiel in their seminal work of 1970~\cite{fama1970} the assertion of the efficient-market hypothesis (EMH) that `{\it a market in which prices always fully reflect available information is called efficient}' and establishing three different levels thereof. In spite of criticisms over its actual validity -- especially its `strong' case --, the EMH is at the hard-core of standard risk-adjustment in mathematical and quantitative finance expressed by means of the martingale and non-arbitrage properties formally introduced by Samuelson~\cite{samuelson1965}, which have been used back in the 1900s by Bachelier and in some sense by Einstein in his theory of Brownian motion~\cite{voit2003statistical} though.
Many of the criticisms EMH has faced~\cite{schwager2012market} rely on arguments that precisely set financial markets as a complex system -- and for this reason --  market efficiency has attracted the attention of complexity scientists, especially physicists~\cite{ausloos2016usual}.

In 2009,  a milestone event occurred in finance with the introduction of the first decentralized cryptocurrency, the bitcoin (BTC).  Since 2011, we have seen a proliferation of alt-coins. According to the UK's Financial Conduct Authority~\cite{FCA2022}, as of 2023, more than twenty thousand cryptocurrencies (coins and tokens) have been created\footnote{While cryptocurrencies coins have their own independent blockchain and play the role of a currency, tokens are built on already existing blockchain and aim to define the worth of a given asset and enhance transactions on the blockchain they have been defined.}.  Such a cornucopia of cryptocurrencies implies that many of them are very little traded, which brings forward questions over liquidity and hence volatility as illiquid cryptos are more sensitive to whale-wallet moves. Accordingly, with the purpose of trading, although penny-cryptos are more appealing -- because the minimal change of its quote (one tick) corresponds to a large relative fluctuation --, there is the likely risk of illiquidity in the event of having to stop the position out and experience an instance equivalent to a squeeze, which is a paradigmatic example of inefficiency. 

From a characterization perspective, the matter of illiquidity -- and by association inefficiency -- is a major ingredient in the degree of complexity of the price fluctuation dynamics. Therefore, the assessment of the latter has been (partly) used as a proxy for the former~\cite{zunino2010complexity,chakraborti2011econophysics,huber2016can}. As detailed in the following section, single entropic measures were applied to cryptocurrencies willing to sort them with respect to their inefficiency degree. Nonetheless, complexity is a sufficiently intricate concept on its own to be properly described by means of a single measure~\cite{lloyd2001measures}, i.e., a unidimensional quantity. For this reason, multivariate measures, namely those combining different sorts of entropy started being used to classify a system within a Complexity Science framework~\cite{rosso2007distinguishing}.

In this paper, we study the relation between complexity and inefficiency of cryptocurrencies using the Binary Complexity-Entropy Plane (BiCEP). Our data is composed of the cryptocurrencies presented in Table~\ref{crypto-ticker-name}. This choice was adopted so that all of the 47 cryptocurrencies we analyzed had the same number of data values. 
The market-cap of this set sums up circa $90\%$ of the total cryptocurrencies worth as of May 2024. The rise/drop in the daily closing price is then associated with 1/0 values. Although a binary description might seem reductionist at first, it is worth noting that the up/down dynamics is used in the definition of relevant technical indicators, namely the On Balance Volume (OBV)~\cite{pring2014technical}, used by many traders -- both retail and institutional -- in their decision-making process\footnote{In the case of OBV instead of 0 the price drop translates into $-1$.} or toy-models on financial transactions~\cite{queiros2007multi}. From a wider point of view, binary sequences have served as a fundamental framework for several areas including biology~\cite{pires2022randomness,macia2022aperiodic} and physics~\cite{dal2012deterministic,pires2020aperiodic,pires2021negative}.

We organize the following sections as follows: In Sec.~\ref{sec:review}, we contextualize our work with respect to the application of block-based measures, the definition of complexity-entropy planes and previous analyses using entropic measures to describe the efficiency of cryptocurrencies. In Sec.~\ref{sec:methods}, we convey the methods we utilize, namely how we compute the components of each crypto in the Binary Complexity-Entropy Plane (BiCEP for short) and define the inefficiency ranking used to sort our set. In Sec.~\ref{sec:results}, we present our results and, last of all, we offer our final remarks in  Sec.~\ref{sec:remarks}.

\begin{table}[t]
\centering
\caption{Ticker and name of the 47 cryptos we have studied in our work.}
\begin{tabular}{l c c c}
\hline
Ticker & Name \\ 
\hline \hline
AAVE & Aave \\
ADA & Cardano \\
ALGO & Algorand \\
ATOM & Cosmos Hub \\
AVAX & Avalanche \\
BCH & Bitcoin Cash \\
BGB & Bitget \\
BNB & Binance Coin \\
BSV & Bitcoin Satoshi's Vision \\
BTC & Bitcoin \\
CRO & Cronos \\
DOGE & Dogecoin \\
DOT & Polkadot \\
EOS & EOS \\
ETC & Ethereum Classic \\
ETH & Ethereum \\
\hline
\end{tabular}
\hspace{0.4cm}
\begin{tabular}{l c c c}
\hline
Ticker & Name \\ 
\hline \hline
FET & Fetch.AI \\
FIL & Filecoin \\
FIRO & Firo \\
FLOW & Flow \\
FTM & Fantom \\
HBAR & Hedera Hashgraph \\
ICP & Internet Computer Protocol \\
INJ & Injective \\
JASMY & JasmyCoin \\
LDO & Lido DAO \\
LEO & UNUS SED LEO \\
LINK & Chainlink \\
LTC & Litecoin \\
MATIC & Polygon \\
MKR & Maker \\
NMC & Mincoin \\
\hline
\end{tabular}
\hspace{0.4cm}
\begin{tabular}{l c c c}
\hline
Ticker & Name\\ 
\hline \hline
NEAR & Near \\
NEO & Antshares \\
OKB & OKB \\
XPM & Primecoin \\
RUNE & THORChain \\
SHIB & Shiba Inu \\
SOL & Solana \\
XLM & Lumen \\
VET & VeChain \\
VTC & Vertcoin \\
XMR & Monero \\
XNO & Nano \\
XRP & Ripple \\
XVG & Verge \\
ZEC & Zcash \\
\hline
\end{tabular}
\label{crypto-ticker-name}
\end{table}

\section{\label{sec:review} Literature review}

In this section, we give a brief overview of the relevant related works.

\subsection{Entropy and complexity}

The Block entropy $H_m$, also known as $N$-gram entropy, is simply the standard Shannon entropy applied to the $k$-history time series of a time series~\cite{shannon1948mathematical}. The $k$-histories are defined by the block size $m$ and the block entropy reduces to the traditional Shannon entropy for $m = 1$. Thus, this measure is commonly used by taking $h = \lim_{m\to\infty} H_m/m$ (Entropy Rate) as the average amount of randomness per symbol that persists after all the correlations and the constraints are taken into consideration~\cite{schurmann1996entropy,lesne2009entropy,Brigatti2021,papadimitriou2010entropy,zapart2009entropy}.
Within this context, the Block Entropy has been applied to time series from the most diverse fields of study, viz., quantitative linguistics~\cite {papadimitriou2010entropy}, foreign exchange~\cite{zapart2009entropy}, atmospheric and space science~\cite{balasis2009investigating}, climate~\cite{larson2011block}
among others.

It should be noted that some of the authors showed some difficulties in calculating the Entropy Rate~\cite{salgado2021estimating}: (a) reaching the limit of infinite block length is impossible in practice; (b) estimating the Block Entropy Rate from empirical measurements requires a large amount of data, even for moderately large block lengths; (c) the finiteness of the observed trajectories can lead to errors that can be associated with censored samples of the Entropy Rate.
In our study, we present an alternative way to use the Block Entropy without the need to calculate the Entropy Rate. Our proposal is supported by two new measures, Block Complexity and Block Disequilibrium, which are jointly used herein to study cryptocurrency efficiency.

\textcolor{black}{
As far as we are aware, the Complexity-Entropy Plane (CEP) was initially proposed in Ref.~\cite{crutchfield1989inferring}, but remains a subject of ongoing discussion~\cite{grassberger2017comment,grassberger2017some}.
Alternative CEP formulations are available \cite{lopez1995statistical,feldman1998measures,rosso2007distinguishing}.
CEPs based on ordinal symbolization have been successfully applied to the analysis of time series  of diverse fields of study in recent years~\cite{ribeiro2012complexity, fernandes2020taxonomy, ma2022complexity, li2023refined, mastroeni2024wavelet, stosic2024generalized, lima2024characterization}. However, the ordinal symbolization is not suitable for time series that have equal values~\cite{zunino2017permutation,cuesta2018patterns}, limiting its application to binary sequences.
}

\subsection{Efficiency of cryptocurrencies}

With respect to efficiency, we can observe diverse and contrasting results in the literature:

\begin{enumerate}

\item The empirical analysis in Ref.~\cite{urquhart2016inefficiency} indicates Bitcoin returns exhibit significant inefficiencies over the full sample period. However, when the sample is partitioned into two subsamples, certain statistical tests reveal evidence of efficiency in the later subsample.

\item The study in Ref.~\cite{nadarajah2017inefficiency} applies a power transformation to the Bitcoin return series. In this case, Bitcoin satisfies the efficiency hypothesis for eight distinct statistical tests. 

\item The author in Ref.~\cite{kristoufek2018bitcoin} provides  evidences that both Bitcoin markets remained predominantly inefficient from 2010 to 2017, with temporary deviations towards efficiency observed during periods following substantial price corrections.

\item Reference~\cite{zargar2019informational} re-examines Bitcoin informational efficiency by analysing price data at multiple time frames, including 15, 30, 60, and 120 minutes, as well as daily intervals. This work reveals informational inefficiency is more pronounced at higher frequencies. 

\item The authors in Ref.~\cite{aggarwal2019bitcoins}  gauge the efficiency of daily Bitcoin returns from July 2010 to March 2018. This study provides evidences against the random walk hypothesis.

\item Reference~\cite{amirat2021exploring} presents a statistical examination of the randomness of cryptocurrency returns; its authors conclude that several cryptocurrencies exhibit significant non-randomness with the exception of Bitcoin. 

\item At odds with the aforementioned work, in Ref.~\cite{fraz2021non} the authors provide results that reject the null hypothesis of weak-efficiency for Bitcoin; it also suggests that it does not exhibit characteristics of a random walk. 

\item \textcolor{black}{On the other hand, the study presented in  Ref.~\cite{puoti2025quantifying} indicates that LTC, BNB, BTC, ETH, and XRP {\it `exhibit characteristics closely resembling those of Brownian noise when analyzed in a univariate context'}. In addition, they understood that simple statistical models `consistently outperform the more complex machine and deep learning' forecast approaches. This hints latter methods tend to overfit data and introduce specious signal elements, as regularly occurs when they treat Brownian motion}.

\item \textcolor{black}{The analysis considering an intraday time frame of other cryptos besides Bitcoin in Ref.~\cite{bariviera2018analysis} showed a varied behavior. The authors found `some kind of persistent stochastic dynamics with Hurst exponents between $0.5$ and $0.7$', which points to moderate nonefficiency, namely for ETH and ETC, which present a more persistent behavior than the others cryptos  analyzed. This agreed with the respective smaller values of permutation entropies they computed for those two cryptos.}

\item \textcolor{black}{Last, in Ref.~\cite{bariviera2018analysis} its authors considered the financial concepts of stability, independence, and resilience  -- which are intimately related to efficiency -- to analyse major stablecoins (Tether, USD Coin, and Binance USD). They assert that these cryptos cannot be considered as monetary anchors because they fail to display superior performance in all three criteria.}
\end{enumerate}

In other words, the matter of efficiency in the crypto market is still under strong debate due to clashing results, as previously pointed out in the literature~\cite{lengyel2021bitcoin}.

\section{\label{sec:methods} Methodology}

\subsection{The Binary Complexity -- Entropy Plane, BiCEP}

Given a probability distribution $p_i$ of block patterns of a binary sequence of size $m$, as further discussed 
in~\ref{sec:appendixA}. We can calculate the Shannon entropy of the block patterns, also known as Block Entropy~\cite{schurmann1996entropy,lesne2009entropy,balasis2009investigating,larson2011block} as
\begin{equation}\label{eq:permutation_entropy}
    H(P) \equiv -\sum_{i = 1}^{W} p_i\ln p_i\,,
\end{equation}

The maximal value of $H$ is obtained when all blocks are equiprobable, i.e., the distribution of 0s and 1s is the same (within a block), $H_{\rm max} = \ln{W} = m \ln 2$. From this, we define the  Normalized Block Entropy $E(P)$, and the Normalized Block Statistical Complexity $C(P)$ as
\begin{equation}\label{eq:normalised_pe}
    E(P) = \frac{H(P)}{ H_{\rm max} }\,,
\end{equation}

\begin{equation}~\label{eq:statistical_complexity}
    C(P) = \frac{D(P,U)E(P)}{D^{\rm max}}\,,
\end{equation}
where $U \equiv 1/W = 2^{-m}$ is the uniform distribution of the block patterns,
\begin{equation}
    D(P,U) = H\left(\frac{P + U}{2}\right) -  \frac{ H(P) + H(U) }{2}
\end{equation}
is the Jensen-Shannon divergence~\cite{pessa2021ordpy} -- or in this case, the Block Disequilibrium in relation to the uniform distribution --, and 
\begin{equation} 
    D^{\rm max} = -\frac{1}{2}\left(\frac{W+1}{W}\ln(W+1)-2\ln(2W)+\ln{W}\right)
\end{equation}
is a normalization constant, which represents the maximum possible value of $D(P,U)$ that takes places when $P = \{ \delta_{1,i} \}_{i = 1,\dots,W}$, where
$\delta_{ij}$ is the Kronecker delta function.  The BiCEP is a bidimensional space that brings together a pair of values for each time series~\cite{Zanin2021ordpatterns}. 

\textcolor{black}{
Resorting to Figure~2 in Ref.~\cite{Zanin2021ordpatterns}, this type of characterization is capable of placing different kinds of dynamics in different regions of the Complexity-Entropy space. For instance, chaotic dynamics presents high levels of complexity and intermediate values of entropy, whereas stochastic colored dynamics will present a lesser degree of complexity. On the extremes, we have deterministic systems like regular oscillations and pure white noise for which the complexity vanishes which minimal and maximal entropy, respectively.
}

Differently from Normalized Block Entropy $E(P)$, the Block Statistical Complexity $C(P)$ must be zero in both extreme cases: fully ordered (when only one pattern occurs) and totally disorder (when all patterns are equally likely to happen). This means the value of $C(P)$ quantifies structural complexity and provides additional information that is not carried by the value of $E(P)$. Furthermore, $C(P)$ is a nontrivial function of $E(P)$ in the sense that for a given value of $E(P)$, there exists a bounded range of possible values for $C(P)$. This happens because $E(P)$ and $D(P)$ are expressed by different sums and there is thus no reason for assuming a univocal relation between $E(P)$ and $C(P)$~\cite{pessa2021ordpy}, as also depicted in Fig.~\ref{fig:fig_edc_m5}.
We would like to emphasize this is a nutshell presentation of the BiCEP. A thorough introduction of this complexity classification and comparison with other methodologies is to be published elsewhere.

Concerning the relation between efficiency and the BiCEP analysis, we consider a Euclidean measure of the distance from the location of a given crypto, $\mathcal{I}$, in the BiCEP to the location of efficiency in the same plane, i.e., $C_{\rm{eff}}= 0$ and $E_{\rm{eff}} =1$. This defines our inefficiency score,
\begin{equation}
\mathcal{I} \equiv \sqrt{(C - C_{\rm{eff}})^2 + (E - E_{\rm{eff}})^2},
\label{efficiency-def}
\end{equation}
from which we rank the cryptos. Taking into consideration the non-negative nature of $E$ and $C$ and also the geometric nature of price evolution, an efficiency measure based on the geometric mean would be appropriate as well.

\subsection{Significance tests}

Taking into consideration a first principles description of the dynamics of crypto assets is not possible, we must support our assertions by statistical significance testing.
To that end, we performed complexity and entropy tests, which are close to standard null-hypothesis testing in statistics (more details in \ref{sec:appendixA}). Namely, for each crypto, we generated a set of new series by shuffling the elements of binary series defined by the data; these sets allow defining minimal and maximal values of the complexity and entropy measures expected for each crypto were it be perfectly efficient during the time span. 
In addition, we employed statistical testing of randomness and pseudorandomness developed within the context of cryptography~\cite{rukhin2001statistical}.

\section{\label{sec:results} Results}

\begin{figure}[htb]
    \centering
    \includegraphics[scale=0.65]{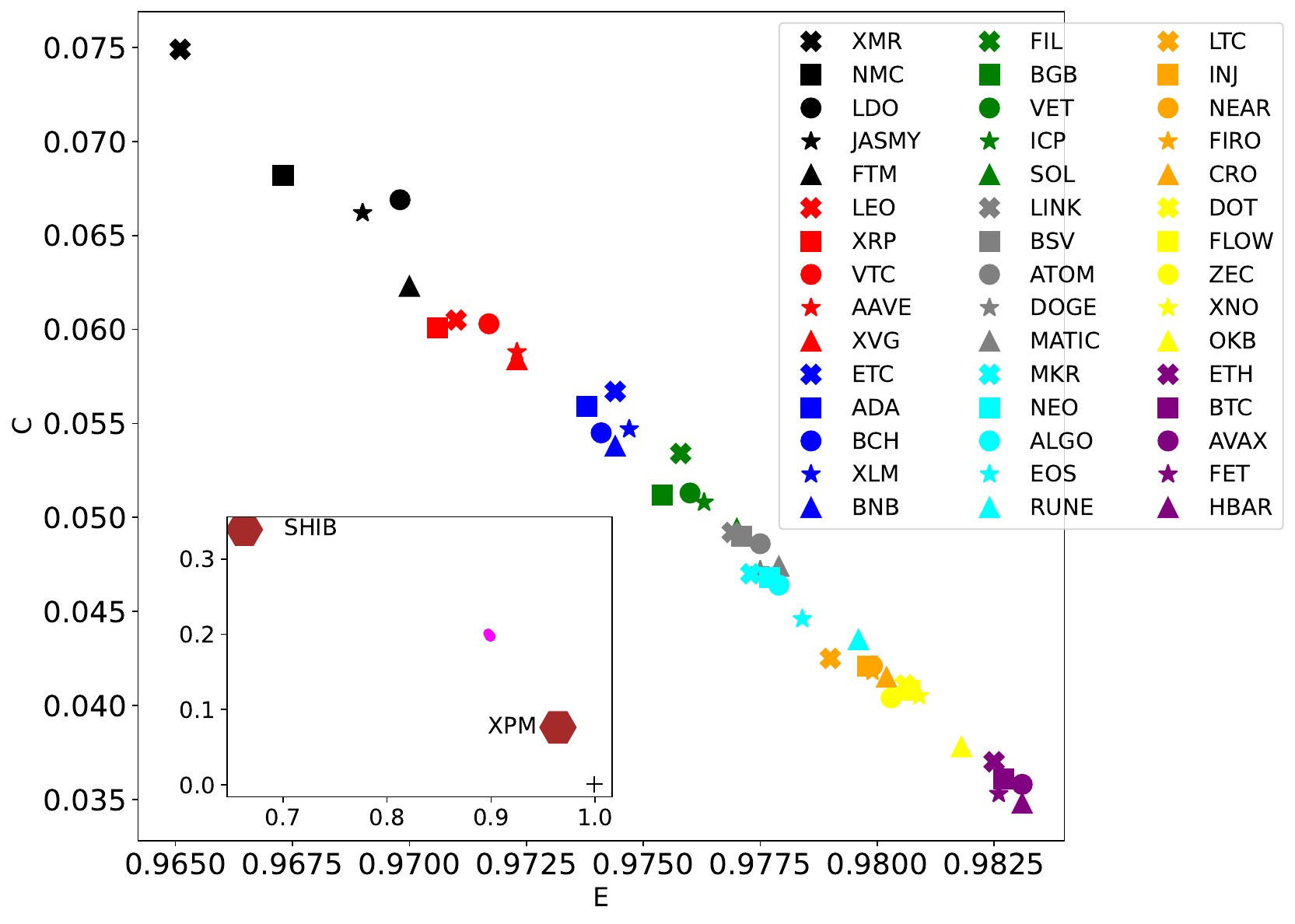}
    \caption{The Binary Complexity-entropy Plane (BiCEP) for the set of cryptos in Table~\ref{crypto-ticker-name}.  The data were collected during the analysis period of August 1, 2021 to November 1, 2024. We used the optimal $m=8$ as delineated in \ref{sec:appendixA}. \textcolor{black}{Within the inset, magenta-colored data points correspond to the mean measures $\{E_s,C_s\}$ obtained from all the 47 shuffled time series, as described in the text. The black plus '+' in the inset correspond to the $\{E_r,C_r\}$ for a purely random binary sequence. } }
    \label{fig:plane_ce_bitflip_m5}
\end{figure}

\begin{table}[h]
\centering
\caption{Location of each crypto in the BiCEP and the respective inefficiency score given by Eq.~(\ref{efficiency-def}) for $m=8$. We have 47 cryptos in this table. \textcolor{black}{To aid in the interpretation of our findings, see the extra results presented in \ref{sec:extra_res}.}  }.
\begin{tabular}{l c c c}
\hline
Crypto & Entropy $(H)$ & Complexity $(C)$ & $\mathcal{I}$ \\ 
\hline \hline
AAVE & 0.972 & 0.059 & 0.065 \\
ADA & 0.974 & 0.056 & 0.062 \\
ALGO & 0.978 & 0.046 & 0.051 \\
ATOM & 0.978 & 0.049 & 0.054 \\
AVAX & 0.983 & 0.036 & 0.040 \\
BCH & 0.974 & 0.054 & 0.060 \\
BGB & 0.975 & 0.051 & 0.057 \\
BNB & 0.974 & 0.054 & 0.060 \\
BSV & 0.977 & 0.049 & 0.054 \\
BTC & 0.983 & 0.036 & 0.040 \\
CRO & 0.980 & 0.042 & 0.046 \\
DOGE & 0.978 & 0.047 & 0.052 \\
DOT & 0.981 & 0.041 & 0.045 \\
EOS & 0.978 & 0.045 & 0.050 \\
ETC & 0.974 & 0.057 & 0.062 \\
ETH & 0.983 & 0.037 & 0.041 \\
FET & 0.983 & 0.035 & 0.039 \\
FIL & 0.976 & 0.053 & 0.059 \\
FIRO & 0.980 & 0.042 & 0.046 \\
FLOW & 0.981 & 0.041 & 0.045 \\
FTM & 0.970 & 0.062 & 0.069 \\
HBAR & 0.983 & 0.035 & 0.039 \\
ICP & 0.976 & 0.051 & 0.056 \\
\hline
\end{tabular}
\hspace{0.1cm}
\begin{tabular}{l c c c}
\hline
Crypto & Entropy $(H)$ & Complexity $(C)$& $\mathcal{I}$ \\ 
\hline \hline
INJ & 0.980 & 0.042 & 0.047 \\
JASMY & 0.969 & 0.066 & 0.073 \\
LDO & 0.970 & 0.067 & 0.073 \\
LEO & 0.971 & 0.060 & 0.067 \\
LINK & 0.977 & 0.049 & 0.054 \\
LTC & 0.979 & 0.043 & 0.047 \\
MATIC & 0.978 & 0.047 & 0.052 \\
MKR & 0.977 & 0.047 & 0.052 \\
NEAR & 0.980 & 0.042 & 0.047 \\
NEO & 0.978 & 0.047 & 0.052 \\
NMC & 0.967 & 0.068 & 0.076 \\
OKB & 0.982 & 0.038 & 0.042 \\
RUNE & 0.980 & 0.043 & 0.048 \\
SHIB & 0.663 & 0.339 & 0.478 \\
SOL & 0.977 & 0.049 & 0.054 \\
VET & 0.976 & 0.051 & 0.057 \\
VTC & 0.972 & 0.060 & 0.067 \\
XLM & 0.975 & 0.055 & 0.060 \\
XMR & 0.965 & 0.075 & 0.083 \\
XNO & 0.981 & 0.041 & 0.045 \\
XPM & 0.964 & 0.076 & 0.084 \\
XRP & 0.971 & 0.060 & 0.067 \\
XVG & 0.972 & 0.058 & 0.065 \\
ZEC & 0.980 & 0.040 & 0.045 \\
\hline
\end{tabular}
\label{crypto-eff-tab}
\end{table}

Applying the methods described in the previous section, we obtained the BiCEP as presented in 
Fig.~\ref{fig:plane_ce_bitflip_m5}  and the inefficiency score, $\mathcal I$ as presented in Tab~\ref{crypto-eff-tab}.
Although a hasty analysis of the plot induces one to assume a trivial linear dependence between Complexity and Entropy, it must be taken into consideration that the results concentrate in a very tiny part of the BiCEP; in terms of the entropy range we are actually talking about $1.8\%$ of the full range. 

To evaluate the performance of our approach, we applied it to series with known levels of randomness. The results, presented in \ref{sec:appendixA}, demonstrate the robustness of BiCEP in characterizing randomness across a variety of controlled conditions.

\textcolor{black}{
From our efficiency significance tests, we found that SHIB failed both entropy and complexity efficiency tests and thus we can classify it as inefficient according to complexity and information measures;  all the other cryptos cannot be classified as inefficient with statistical significance.
Explicitly, the null hypothesis was discarded in both tests for SHIB, indicating that the order of the data does not occur at random.
}

Analysing the inefficiency score provided by $\mathcal I$, we verify that there is coherence between the values in Table~\ref{crypto-eff-tab} and the significance test since the cryptos exhibiting larger values of the efficiency distance are those which failed the efficiency significance test. 
Applying the randomness and pseudorandomness test, we substantiated the inefficiency of SHIB and the non-inefficiency of our $\mathcal I$ Top~5; in this case, ETH is at the limit of significance when the $p$-value is equal to $5\times 10^{-3}$ though. For this test, the best result is for FLOW followed by OKB whereas considering $\mathcal{I}$ score Top~5 the best stand is for AVAX in 4th place. 

In respect of the less inefficient ranking established from the $\mathcal I$ scores, the first tier is composed of HBAR, AVAX, FET, BTC, ETH, which appear rather isolated from the rest. Recall the smaller the inefficiency score, the higher the efficiency. 
We shall center our attention on this set of five cryptos; looking at the canonical elements alluding to efficiency of an asset in a financial market such as market capitalization or trading volume, we observed some relation between the $\mathcal{I}$-based crypto ranking and efficiency insofar as both BTC and ETH are found within the top tier. In addition,  we followed this up with an analysis  of the Kendall correlation results between the ranking given by $\mathcal{I}$ and those established by sorting market-cap and trading volumes, respectively. Nonetheless, those two cornerstone cryptos are not the first and runner-up alt-coins in terms of efficiency.

Let us try to shed light on the difference between cryptos and financial assets like stocks, fiat money, and commodities. To that, we start surveying the main crypto, the Bitcoin, which accounts for as much as $54\%$ of the total crypto market-cap. The Non-Turing complete nature of BTC  hampers its functionality  which provides robust security to it; however, this advantage comes at the cost of the other features that blockchain technology tries to optimize, namely its scalability and capacity to decentralize due to the Proof of Work~\cite{proofofwork} consensus mechanism assumed.
On the other hand, the second largest market-cap crypto -- i.e., ETH -- is Turing complete; it uses a Proof of Stake~\cite{arslanian2022ethereum}
consensus mechanism and supports a smart contracts approach~\footnote{Smart contracts are digital contracts stored on a blockchain that are automatically executed when predetermined terms and conditions are met.}. This makes ETH a quite scalable and decentralized coin at the cost of security as the Proof of Stake consensus mechanism is more susceptible to malware and network failure.
On the other hand, we have lesser traded cryptos -- namely, HBAR, AVAX, and FET --  that have shown little inefficiency outperforming the other two.

Taking into consideration that FET is a token of ETH, we understand that its higher liquidity arises from the fact that this token is the only means of value exchange on fetch.ai, which is an open-access decentralized machine learning network that stems from blockchain technology.
To further investigate the connection between FET price dynamics and artificial intelligence (AI), we computed the correlation between the price variations  of FET and the Global X Robotics \& Artificial Intelligence ETF (BOTZ)~\cite{botz}.
As shown in \ref{sec:appendixFET}, in spite of the spell after the announcement that fetch.ai had entered administration during 2024Q1~\cite{ccn-fet}, the qualitative (up/down) price fluctuations of this paradigmatic AI coin shows a strong correlation with both BOTZ and -- naturally -- BTC.

Considering non-tokens, Avalanche was created in order to present the highest scalability, decentralization and security in the world by making use of a combination of smart contract technology and a validation system that blends the best of the Proof of Work and Proof of Stake consensus. For instance, the Avalanche Consensus is able to process in excess of 4500 transactions per second with little latency and minimal costs, whereas ETH manages to make around 30 transactions per second. Naturally, such a high-frequency flux of trading translates into the trading volume in the daily time frame. The AVAX white paper can be consulted for further details~\cite{AvalanchePlatform}.

Last -- but absolutely not least -- we have Hedera~\cite{hedera}, which is the only public ledger that uses hashgraph instead of the mainstream blockchain consensus mechanisms.\footnote{Still, the consensus relies on a variant of the Proof of Stake.} 
While BTC and ETH perform around 10 and 30 transactions per second, respectively, HBAR manages to process in excess of 10 thousand transactions each taking around 5 seconds to get confirmed whereas ETH trebles that time-span and BTC has got a transaction confirmation time larger than 10 minutes~\cite{btc-eth-hbar}.
The advantages are also environmental because the energy consumption per HBAR transaction is infinitesimal in comparison with BTC ($2\times 10^{-5} \%$) and 25 times less power consuming than the already parsimonious AVAX.

On the other extreme, despite the fact that SHIB is based on a Proof of Stake like ETH, the token is highly concentrated on a few whale-wallets~\cite{insights}. At first, this could be considered beneficial since in owning a larger stake of the crypto one has a higher chance of getting to add another block; however, for a transaction to occur one needs a buyer and a seller -- which is affected by concentration -- a preponderant fact in liquidity and at the end of the day in the efficiency of the asset.

\section{\label{sec:remarks} Final Remarks}

In this paper, we analyzed  the central bulk of the non-stablecoin cryptocurrency ecosystem with the goal of classifying them according to its efficiency in a daily time frame. To that end, we set forth a Binary Complexity-Entropy Plane --  BiCEP -- to be formally introduced elsewhere, which allows us to study the efficiency of the crypto considering two measurements: Entropy and Complexity. The maximal efficiency is described by Entropy equal to $1$ and Complexity equal to $0$ corresponding to a memoryless and random series. 

The mapping of the price dynamics into the simplest rise/drop dynamics permitted us to focus on the key elements of random walk evolution that have been considered the bedrock of the analytical description of the Efficient Market Hypothesis. This assertion of ours follows the same reasoning that `symbolization can increase the efficiency of finding and quantifying information from' complex `systems, reduce sensitivity to measurement noise, and discriminate both specific and general classes of proposed models'~\cite{hirata2023review} and particularly adapts to the matter of efficiency.

Afterwards, the cryptos were ranked according to the inefficiency score of each one, which corresponds to the distance from the crypto location on the BiCEP to the maximal efficiency point. To support our analysis, we have designed significance tests for complexity and entropy.

\textcolor{black}{
Having in mind that the price fluctuations can be considered as the composition of the direction of motion (sign of the return)  and its amplitude -- which is directly related to volatility -- a future analysis on the complexity of volatility in cryptos can furnish a wider picture of their price fluctuations complexity as a whole, particularly in respect of the effective differences between considering the BiCEP approach and the framework based on permutation entropy.
}
\textcolor{black}{
On the other hand, it is possible to explore the local fluctuations around the long-term behavior we surveyed following the lines assumed in other quantitative finance studies~\cite{sigaki2019clustering,graczyk2017intraday,chakraborty2024dynamic} and by establishing in this case a local inefficiency score.
}

From the set of $47$ cryptos we have investigated totalling as much as $90\%$ of the total worth of the crypto market (including stablecoins), SHIB is significantly inefficient. All the remaining 46 cryptos of our set cannot be classified as inefficient in statistical terms. The robustness of our claims was positively tested against modifications in the free parameters of the method, namely the block size and time span.
Therefrom, we conclude that the largest stake of crypto trading occurs to close-to-efficient conditions that we understand as semi-strong efficiency following Famma's classification scheme. 
From a straight quantitative finance perspective, further similar approaches to what we provided an account of can be carried out, namely for the volatility, trading volume  -- after a proper symbolization -- and aggression of the order book.

When we look at the inefficiency score an interesting trait of the crypto-market emerges. The cryptos with smaller values of the inefficiency score are not the cases with larger market-cap or trading volume. Actually, the podium of the least inefficient crypto of our dataset, namely, Hedera (HBAR), Fetch (FET), and Avalanche (AVAX), rank each at $25th$, $19th$, and $8th$ in terms of market capitalization, whereas for trading volume the same cryptos rank at $32nd$, $26th$, and $10th$, respectively. Looking the other way round, the largest and most traded crypto, the Bitcoin (BTC), ranks at $4th$ position regarding its inefficiency score and the runner-up Ethereum (ETH) is $5th$ on the efficiency ranking. This suggests that the factors paving the way to efficiency in crypto-trading go farther afield in comparison to other financial assets like stocks, namely it has likely to do with the features of the crypto asset such as consensus method and validation. Our assertion is corroborated by surveying the characteristics of HBAR, AVAX which were designed to overcome the scalability, decentralization, and security handicaps known by heavily traded cryptos.

Our results also enhance the analysis of the impact of forking: the splitting of a blockchain into two different branches that usually occurs because of further research on a crypto willing to introduce technological advances as well as new functions in it. Looking at BCH, which is a fork of BTC aimed at bringing forth a solution to the Bitcoin scalability problem, we have found that this crypto is twice as inefficient as BTC considering our BiCEP analysis. In  Ref.~\cite{kim2023after}, it has been `conjectured that the efficiency of BCH is lower than BTC'; our results go along with that conjecture by bolstering it with statistical significance. Interestingly, BSV (the fork of the fork) presents a slightly lower $\mathcal I$ score than that of BCH, but still greater than $\mathcal I_{BTC}$. This indicates that, at least for the time being, the solutions set forth by these two fork cryptos have not been effective, probably due to weaker security and the congenital energy/environmental costs imposed by their architectures.

Furthermore, we will implement complementary  methodologies, such as the recently developed lacunarity-persistence plane~\cite{pires2024parrondo}, to characterize the dynamics price movements.

\appendix

\section*{Acknowledgements}
S.M.D.Q. thanks CNPq (Grant No. 302348/2022-0) and FAPERJ (SEI-260003/005741/2024) for financial support.

\section{Details and validation}\label{sec:appendixA}

In this Appendix, we provide details of our methodology as well as our validation tests on binary sequences with controlled levels of randomness.

\subsection{Details}

\begin{figure}[htb]
    \centering
    \includegraphics[scale=0.40]{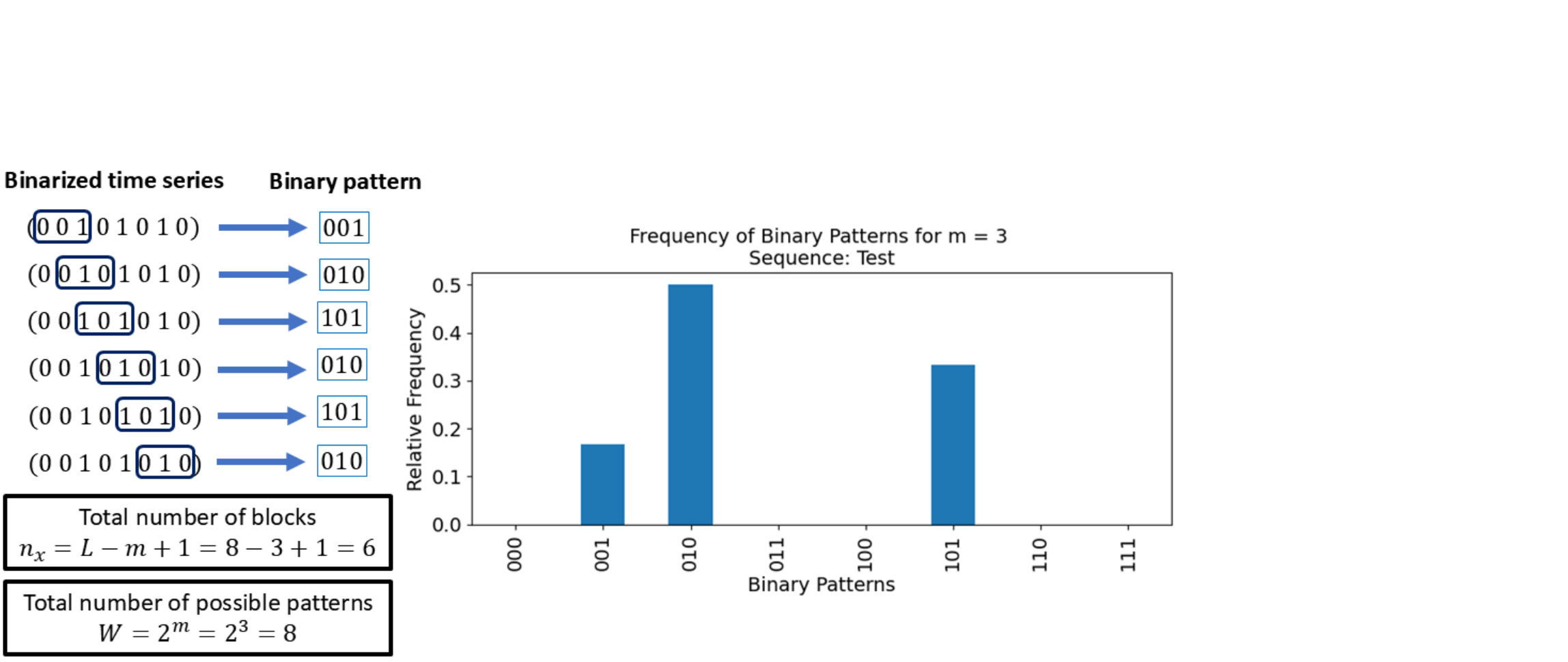}
    \caption{Illustration of the process of symbolizing block patterns for a time series of size $L = 8$ using a block size of $m = 3$.    }
    \label{fig:diagrama-lacunar}
\end{figure}

Figure~\ref{fig:diagrama-lacunar} illustrates several important steps.
We write the time series as a $1 \times L$ array, with $L$ being the size of the series. Thus, each element of the array $A =[a_1j]_{j = 1, \dots, L}$ corresponds to an observation of our time series. The probability distribution denoted as $P = \{p_i\}_{i = 1, \dots, W}$, is essentially the proportional occurrence of all conceivable patterns within the process of symbolizing block patterns.
In other words,
\begin{equation}\label{eq:permutation_probability}
    p_i = \frac{\mathrm{number \; of \; blocks} \; B_i \; \mathrm{with \; type}\; s_i \; \mathrm{in} \; A}{n_x},
\end{equation}
To identify block patterns in an array $A$, the array is partitioned into overlapping sub-arrays $B = [b_{1j}]_{j=1}^{m}$, where each $B$ is a row vector of length $1 \times m$ herein referred to as blocks. The number $m$ of elements in each sub-array is defined as the block size.
The number of possible partitions (or blocks) is given by $n_x = L - m + 1$. Thus, considering the array $A = [B_i]_{i=1}^{L-m+1}$, the order (or absence) in which lacunar points appear in blocks $B_i$ defines the corresponding block pattern, with the total number of possible patterns given by $W = 2^m$. Once this process is complete, the statistics of the resulting block pattern distribution is computed. An example is provided in Fig.~\ref{fig:diagrama-lacunar}.

\subsection{Validation and significance testing}

\begin{figure}[htb]
    \centering
    \includegraphics[scale=0.30]{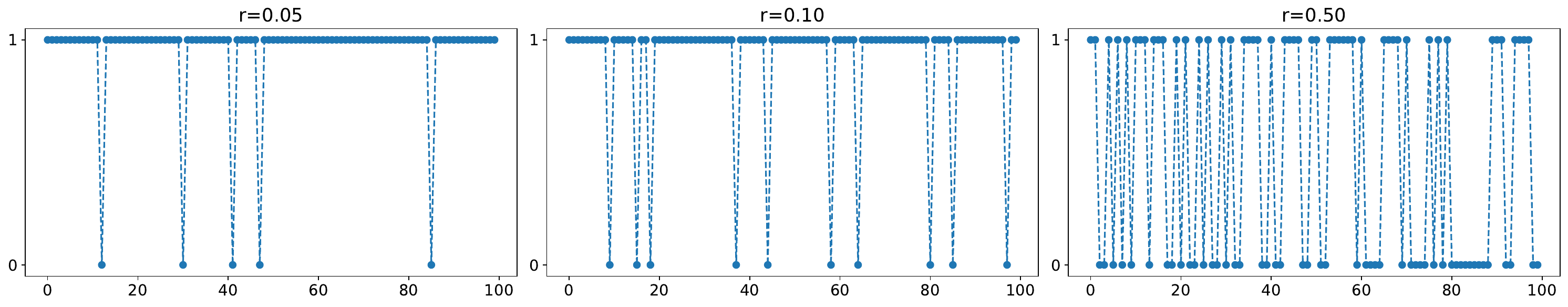}
    \caption{RBF sequences with controlled randomness. }
    \label{fig:bitflip_serie}
\end{figure}

Figure~\ref{fig:bitflip_serie} shows examples of random bit flip (RBF) sequences with the randomness characterized by the parameter \textcolor{black}{$r$}. These panels in the figure illustrate the different ranges of pattern persistence for distinct values of \textcolor{black}{$r$}. To generate an RBF sequence of length $L$, we first create a constant array consisting solely of ones, with \textcolor{black}{$r$} indicating the fraction of bits to be flipped. Next, we randomly select \textcolor{black}{$L_r$} indices within the array where the bit flips will occur. Each index is selected without replacement to ensure that all flips are unique.
At each selected index, the bit is flipped: a bit with value $0$ is changed to $1$, and a bit with value $1$ is changed to $0$. This method allows for the introduction of controlled randomness into binary sequences.

\textcolor{black}{
In order to grant statistical significance to our results we performed as follows: first, we compute the entropy $E_o$ and complexity $C_o$ for the original sequence. Then, $N=10^3$ random surrogates are defined from the original sequence with both the entropy $E_r$ and complexity $C_r$ computed for each case. Finally, $\{E_o,C_o\}$ are checked to see if they are within the range of $\{E_r,C_r\}$. If the original sequence satisfies the previous condition, we say that the null hypothesis cannot be ruled out, i.e., all possible orders of the data are equally likely. 
Complementary, we drawed on a statistical testing of randomness and pseudorandomness developed within the context of cryptography the details of which can be found in Ref.~\cite{rukhin2001statistical}.
}

\subsection{BiCEP results for controlled series}

\begin{figure}[htb]
    \centering
    \includegraphics[scale=0.5]{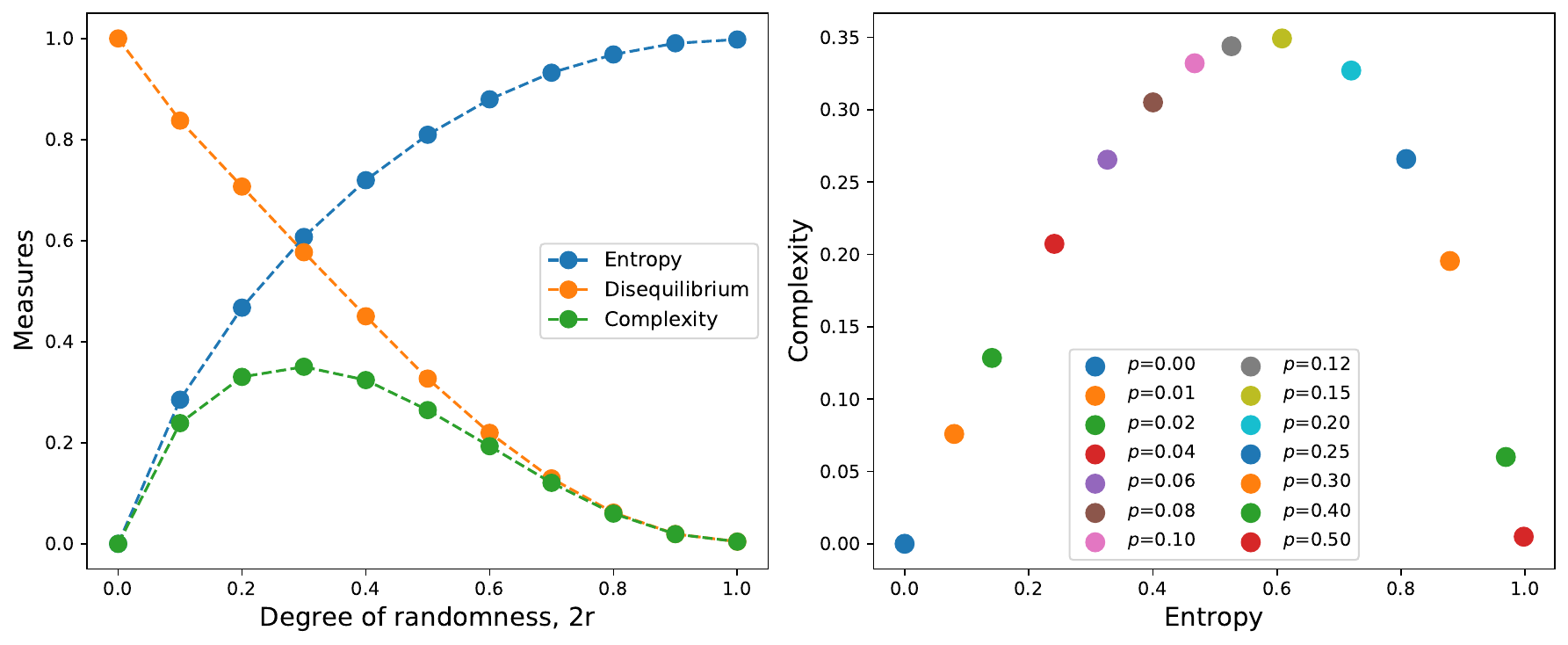}
    \caption{Characterization of RBF time series for block size $m = 8$ with several measures. (a) Entropy $E$, Disequilibrium $D$, and Complexity $C$. (b) Binary Complexity-Entropy plane (BiCEP). }
    \label{fig:fig_edc_m5}
\end{figure}

Figure~\ref{fig:fig_edc_m5} furnishes a comprehensive view of the entropy ($E$), disequilibrium ($D$), and complexity ($C$) for the case of block size $m = 8$ and \textcolor{black}{$0 \leq 2\,r \leq 1$}. 
While $E$ and $D$ exhibit monotonic increasing and decreasing trends with respect to \textcolor{black}{$r$}, the complexity $C$ exhibits a nonmonotonic behavior. This nonmonotonicity highlights the intricate nature of complexity. Indeed, complexity, defined as the product of disequilibrium and entropy, $C = D. E$, reflects the interplay between order and disorder within the sequences.

\subsection{Optimal block size}

\begin{figure}[htb]
    \centering
    \includegraphics[scale=0.55]{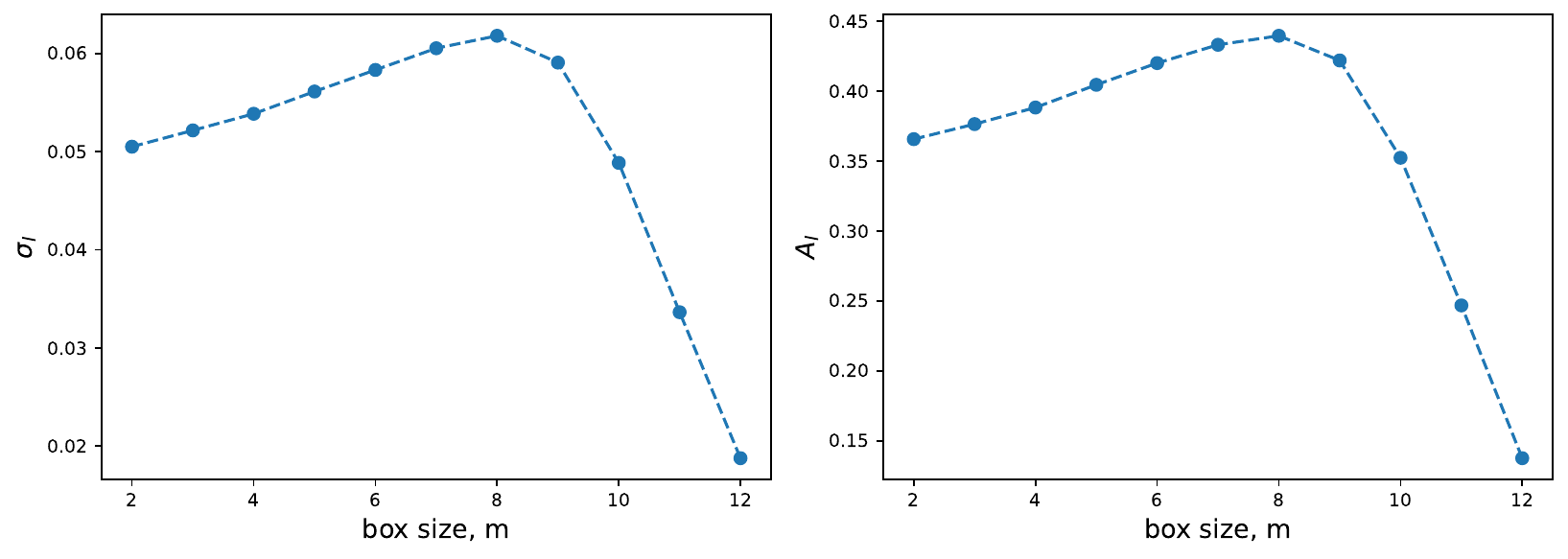}
\caption{Standard deviation (left) and amplitude (right) of all the estimated  inefficiencies 
$\mathcal{I}$ for several $m$ for our dataset of cryptocurrencies. 
Both cases peak at $m=8$, implying this value maximizes the distinguishability in the BiCEP of our dataset.
}
    \label{fig:fig_optimal_m}
\end{figure}

\textcolor{black}{To enhance the distinction between our time series, we must determine the optimal block size.}
\textcolor{black}{
Figure~\ref{fig:fig_optimal_m} depicts the standard deviation and amplitude of the estimated inefficiencies $\mathcal{I}$ as a function of increasing the \textcolor{black}{block} size $m$, considering the set of cryptocurrencies presented in Table~\ref{crypto-ticker-name}. A discernible peak is observed at $m=8$, indicating that this value maximizes the discriminatory power of BiCEP for our dataset.
}

\section{Correlations between FET and BOTZ (BTC) }\label{sec:appendixFET}

\begin{figure}[htb]
    \centering
    \includegraphics[scale=0.35]{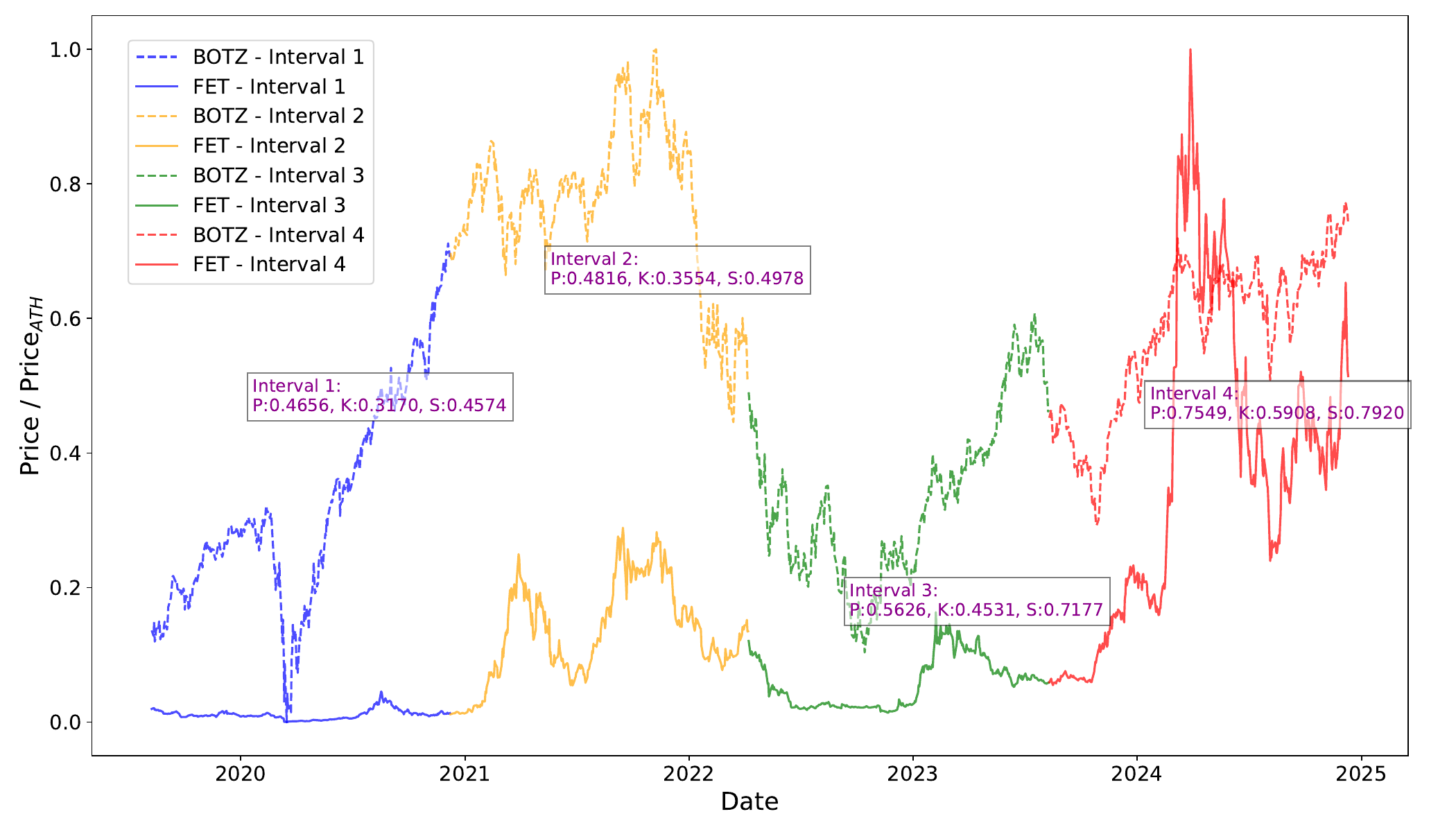}
    \includegraphics[scale=0.35]{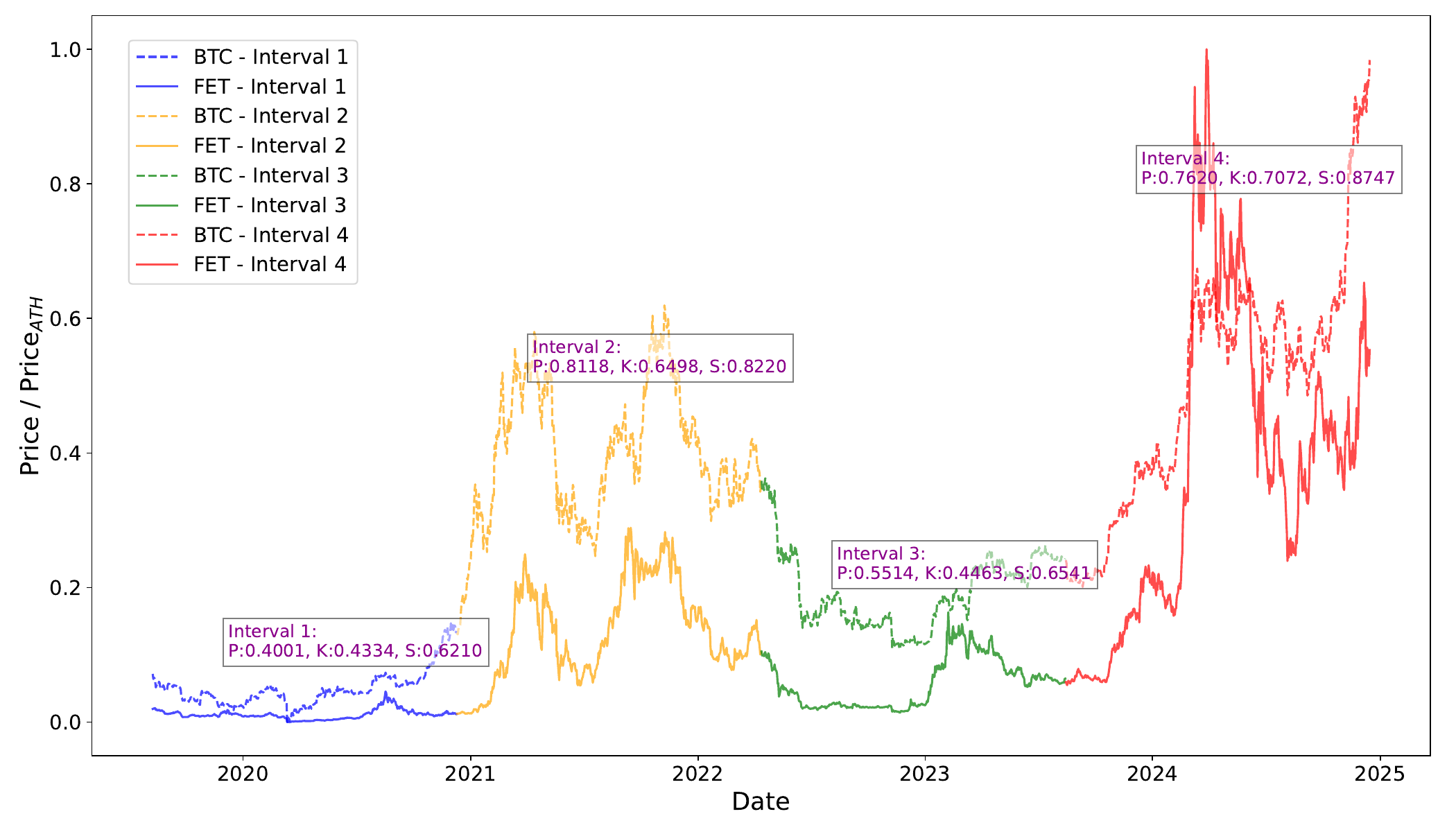}
\caption{Upper panel:
FET and BOTZ exchange traded fund prices v Date. Lower panel: FET and BTC vs Date. 
In both panels the values for each segment -- which is composed of 336 quotes for each series -- are presented in the plot where P,K, and S stand for Pearson, Kendal, and Spearman up/down correlation function, respectively. The values of both instruments were normalized by its all-time high (ATH).
}
\label{fig:cor-fet-botz-btc}
\end{figure}

\textcolor{black}{Herein, we show in the panels of Figs.~\ref{fig:cor-fet-botz-btc} the plots and and results related to the analysis we performed between the up/down fluctuations of FET with the exchange traded fund BOTZ and BTC. Looking at the correlations yearly, we verified that the Pearson correlation between FET and BOTZ is never less than 0.45, which bolsters our assertion that FET would follow the snowballing of AI by nature.}\footnote{Needless to say that FET is not the only payment method for AI. Moreover, we are not discussing causal relations between instruments as well.}

The same happens with the correlations between FET and BTC that reach values as high as $0.81$ for the Pearson correlation in $2021$.
In $2024$, even with the financial problems disclosed by the fetch.ai company and the Bitcoin halving process, the correlations remained quite significant.

\section{Extra results}\label{sec:extra_res}

\begin{figure}[htb]
    \centering
    \includegraphics[scale=0.45]{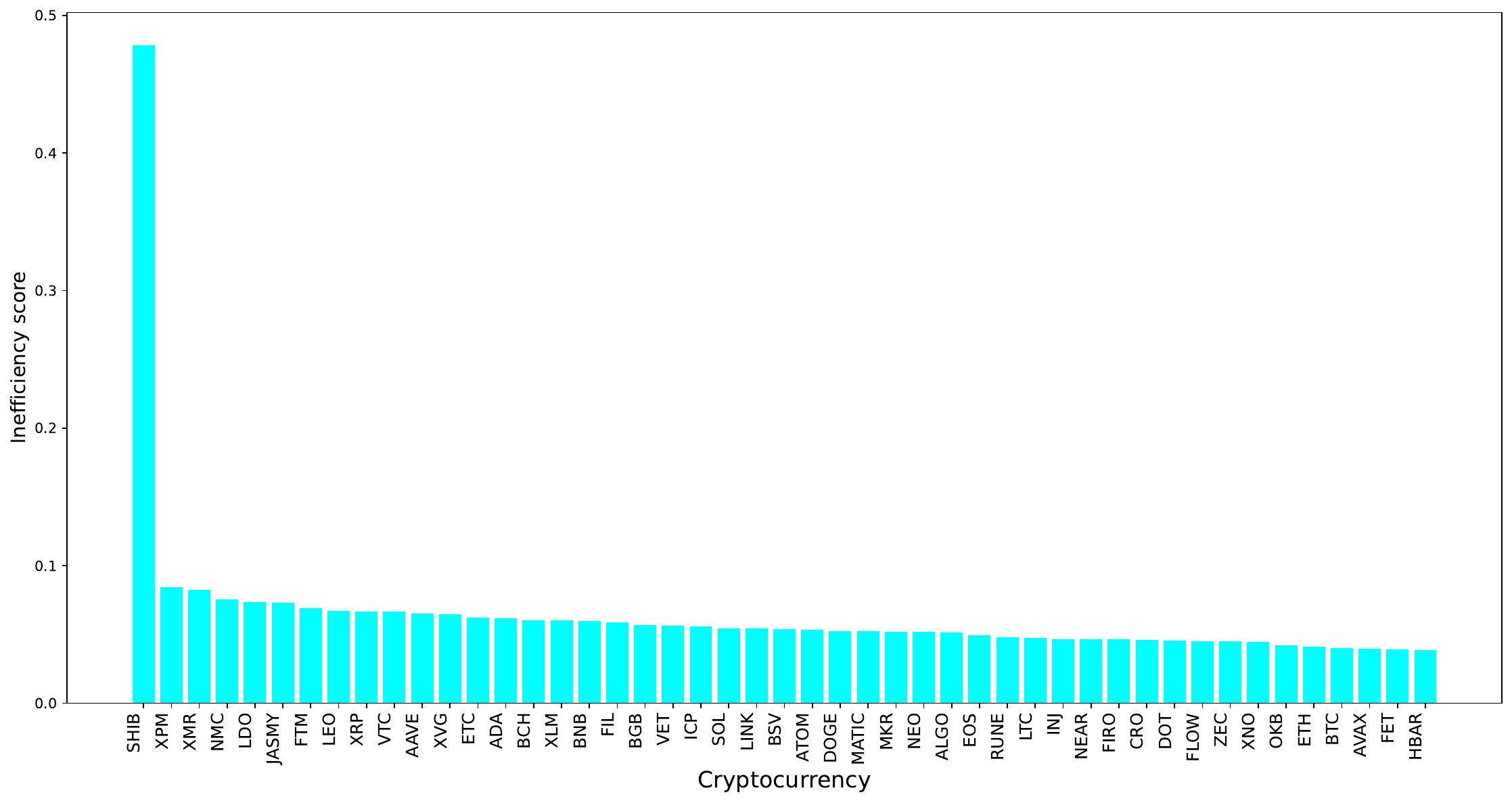}    
\caption{Barplot depicting the inefficiency scores of the analyzed cryptocurrencies. }
\label{fig:barplotIneff}
\end{figure}
 
\begin{figure}[htb]
    \centering
    \includegraphics[scale=0.4]{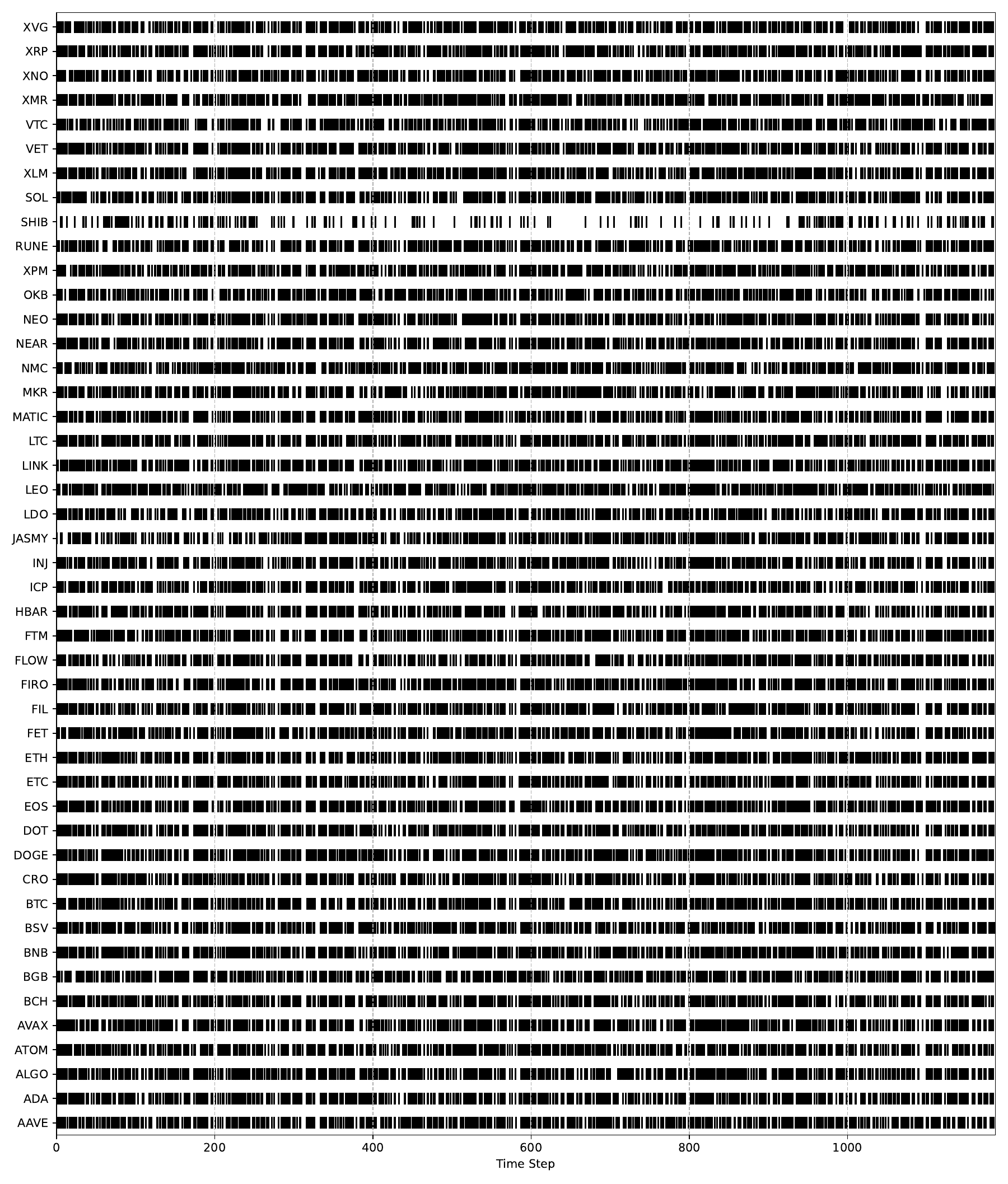}    
\caption{Barcode-like visualization of the binarized temporal sequences generated from the cryptocurrency assets listed in Table~\ref{crypto-ticker-name}. }
\label{fig:barcode}
\end{figure}

\textcolor{black}{Figure~\ref{fig:barplotIneff} illustrates the quantitative assessment of cryptocurrency inefficiency, derived from the BiCEP framework. The analysis indicates that Shiba Inu (SHIB) presents a statistically significant higher inefficiency score compared to the remaining cryptocurrencies in the dataset.}

\textcolor{black}{Figure~\ref{fig:barcode} presents a barcode-like representation of the price motion, where the daily closing price changes are translated into a binary sequence. 
An upward shift in the daily closing price is denoted by '1', whereas '0' represents either a constant or downward trend.
Each horizontal line in the barcode corresponds to a specific cryptocurrency asset with the x-axis representing the progression of time.
The black vertical lines within each row denote the occurrences of '1's, effectively visualizing the temporal pattern of daily price increases. This method provides a clear and intuitive way to observe the underlying patterns of price movements.}

\footnotesize
\setlength{\bibsep}{0pt plus 0.3ex}
\bibliography{refs}

\end{document}